\documentclass[11pt]{cernrep}
\usepackage{graphicx}
\usepackage{here}
\usepackage{amssymb}
\def\alphas{\alpha_s}

\def\gsim{\lower0.5ex\hbox{$\stackrel{>}{\sim}$}}
\def\lsim{\lower0.5ex\hbox{$\stackrel{<}{\sim}$}}
\begin{document}
 \title{Resummation for the Tevatron and LHC boson production at small 
$x$\footnotetext{
Contributed to Workshop on Physics at TeV Colliders, Les Houches, France,
26 May - 6 June 2003.}}
\author{Stefan Berge${}^1$, Pavel Nadolsky${}^1$,  Fredrick Olness${}^1$, 
and C.-P. Yuan${}^2$}
\institute{
${}^1$Southern Methodist University, Department of Physics,
         Dallas, TX 75275-0175, U.S.A.
\\
${}^2$Michigan State University, Dept. of Physics and Astronomy,
           East Lansing, MI 48824-1116, U.S.A.
}
\maketitle
\begin{abstract}
Analysis of small-$x$ semi-inclusive DIS hadroproduction 
suggests that multiple parton radiation leads to a broadening 
of transverse momentum ($q_T$) distributions beyond that predicted
by a straightforward calculation in the Collins-Soper-Sterman formalism.  
This effect can be modeled by a modification of the resummed form 
factor in the small-$x$ region. We discuss the impact of such a modification
on the production of electroweak bosons at hadron-hadron colliders.
We show that if substantial small-$x$ broadening is observed in forward
$Z^0$ boson production in the Tevatron Run-2, it would strongly
affect predicted $q_T$ distributions for $W$, $Z$, and Higgs boson
production at the Large Hadron Collider.
\end{abstract}


In the production of electroweak bosons, precise knowledge of the 
transverse mass $M_T$ and transverse momentum
$q_T$ provides detailed information about the 
production process, including the mass of the boson and associated radiative 
corrections.  At the Tevatron, $q_T$ distributions of $Z^0$ 
bosons offer insight into soft gluon radiation, and this information 
is then used for precision extraction of the $W$ boson mass. 
At the LHC, good knowledge of the transverse distribution 
of Higgs bosons $H^0$ will be needed to efficiently separate Higgs boson 
candidates from the large QCD background. Accurate predictions 
for the small-$q_T$ region are obtained via resummation of large
logarithms $\ln^n(q_T/Q)$ arising from unsuppressed soft and
collinear radiation in higher orders of perturbation theory.

As we move from the 2 TeV Tevatron to the 14 TeV LHC, 
typical values of partonic momentum fractions  $x$ 
for producing $W$, $Z^0$, and $H^0$ bosons  become 
smaller,  thus enhancing $\ln(1/x)$ terms in higher orders of
$\alphas$. It is not entirely known how these terms 
(not included in a fixed-order cross section or conventional 
$q_T$ resummation) will affect $W$, $Z^0$, and $H^0$ production 
at the LHC energies, in part because no Drell-Yan $q_T$ data
is available yet in the relevant region of $x$ 
of a few $10^{-3}$ or less.

Studies \cite{Nadolsky:1999kb,Nadolsky:2000ky} in the crossed channel
of semi-inclusive deep-inelastic scattering (SIDIS) suggest that
hadronic $q_T$ distributions at small $x$ cannot be 
straightforwardly described within the Collins-Soper-Sterman (CSS) resummation 
framework~\cite{Collins:1984kg}, if the nonperturbative Sudakov function
behaves like its large-$x$ counterpart from the Drell-Yan process. 
A $q_T$ distribution in SIDIS at $x < 10^{-2}$ is
substantially broader than the conventional CSS prediction. 
The broadening effect can be modeled by including an extra
$x$-dependent term in the Sudakov exponent. To describe the data, 
the extra term must grow quickly as $x \rightarrow 0$. 
It noticeably contributes to the
resummed form factor at intermediate impact parameters 
($b \sim 1/q_T < 1\mbox{~GeV}^{-1}$),
 which hints at its origin 
from perturbative physics. 
A possible interpretation of this term is that it mimics 
higher-order contributions  of the form $\alphas^m \ln^n (1/x)$, 
which are not included in the resummed cross section. Due to the
two-scale nature of the $q_T$ resummation problem, the non-resummed
$\ln(1/x)$ terms may affect the $q_T$ distribution even when 
they leave no discernible trace in inclusive DIS structure functions. 
The DIS structure functions 
depend on one hard scale 
(of order $Q$),  while the CSS resummation formula (cf. Eq.~(\ref{eq:w})) 
also includes contributions from large impact parameters $b$ (small momentum
scales).  As $b$ becomes large, the series 
$\alphas^m(1/b) \ln^n (1/x)$ in the CSS formula may begin to diverge
at a larger value of $x$ than the series 
$\alphas^m(Q) \ln^n (1/x)$ in the inclusive structure functions. 
For this reason, transition to $k_T$-unordered (BFKL-like \cite{BFKL}) 
physics may happen at larger $x$ in $q_T$ distributions than in
inclusive (one-scale) observables.

The $q_T$ broadening discussed above was observed in semi-inclusive
DIS processes. In this study, we explore its possible 
implications for the (crossed) Drell-Yan process. We begin by
examining the resummed transverse momentum distribution for the Drell-Yan 
process \cite{Collins:1984kg}, following notations from 
Ref.~\cite{Landry:2002ix}:
\begin{equation}
\frac{d\sigma}{dy d q_T^2} 
=
\frac{\sigma_0}{S} \ 
\int
\frac{d^2 b}{(2\pi)^2} \ 
e^{-i \vec{q}_T \cdot \vec{b}} \ 
\widetilde{W}(b,Q,x_A,x_B) +Y(q_T,Q,x_A,x_B).
\label{eq:w}
\end{equation}
Here $x_{A,B} \equiv Q e^{\pm y}/\sqrt{S}$, the integral is  the 
Fourier transform of a resummed form factor $\widetilde W$ given in impact 
parameter ($b$) space, and $Y$ is a regular (finite at $q_T \rightarrow 0$) 
part of the fixed-order cross section. In the small-$b$ limit, 
the form factor $\widetilde{W}$ is given by a product 
of a perturbative Sudakov exponent $e^{-{\cal S}_P}$ and 
generalized parton distributions ${\cal \overline{P}}(x,b)$:
\begin{equation}
\left.\widetilde{W}(b,Q,x_A,x_B)\right|_{b^2 \ll \Lambda_{QCD}^{-2}} 
=
e^{-{\cal S}_P(b,Q)} 
\ {\cal \overline{P}}(x_A, b)  \ {\cal \overline{P}}(x_B,b).  \ 
\label{eq:p}
\end{equation}
At moderately small $x$, where the representation (\ref{eq:p}) for 
$\widetilde W$  holds, we write these generalized parton distributions 
in the form
\begin{equation}
\left.{\cal \overline{P}}(x,b)\right|_{b^2 \ll \Lambda_{QCD}^{-2}} \simeq 
({\cal C}\otimes f)(x,b_0/b) \  e^{- \rho(x) \, b^2},
\label{eq:Psmallx}
\end{equation}
where  ${\cal C}(x,b_0/b)$ are coefficient functions, 
 $f(x,\mu)$ are  conventional parton distributions, 
and $b_0=2e^{-\gamma_E}=1.12...$ is a commonly appearing constant factor. 

The term $e^{- \rho(x) \, b^2}$ in $\overline{\cal P}(x,b)$
will provide an additional $q_T$ broadening,
with an $x$ dependence specified by $\rho(x)$. For example, it may 
approximate $x$-dependent higher-order contributions 
that are not included in the finite-order expression for $({\cal C}\otimes f)$.
We parametrize $\rho(x)$ in the following functional form: 
\begin{equation}
\rho(x) = 
c_0 \left( \sqrt{\frac{1}{x^2}+ \frac{1}{x_0^2}}
-\frac{1}{x_0}\right),
\label{eq:ax}
\end{equation}
such that $\rho(x) \sim c_0/x$ for $x \ll x_0$, and
$\rho(x) \sim 0$ for $x \gg x_0$.
This parameterization ensures that the formalism reduces to the usual 
CSS form  for large $x$ ($x \gg x_0$) and introduces an additional source of
$q_T$ broadening (growing as $1/x$) at small $x$ ($x \ll x_0$). 
The parameter $c_0$ determines 
the magnitude of the broadening for a given $x$, while $x_0$ specifies
the value of $x$ below which the broadening effects become important.
In principle, $c_0$ and $x_0$ may depend on the hard scale $Q$;
in this first study, we neglect this dependence. Based on the observed 
dependence 
$\rho(x) \sim 0.013/x$ at $x \lesssim 10^{-2}$ in SIDIS energy flow
data~\cite{Nadolsky:2000ky}, we choose $c_0=0.013$ and $x_0=0.005$ 
as a representative choice for our plots.

As $x\rightarrow 0$, the additional broadening term 
in Eq.~(\ref{eq:Psmallx}) 
affects the form factor $\widetilde{W}$ both at perturbative 
($b \lesssim 1\mbox{~GeV}^{-1}$) and nonperturbative ($b \gtrsim
1\mbox{~GeV}^{-1}$) impact parameters. In addition, the resummed cross
section contains conventional non-perturbative contributions
from power corrections, which become important at large impact parameters 
($b \gtrsim
1\mbox{~GeV}^{-1}$). 
We introduce these corrections by replacing the impact parameter $b$ in 
functions ${\cal S}_P$ and $({\cal C}\otimes f)$ with a variable 
$b_* = b/\sqrt{1+b^2/(0.25\mbox{~GeV}^{-2})}$~\cite{Collins:1984kg} 
and including 
a nonperturbative Sudakov exponent 
$\exp{\{-{\cal S}_{NP}(b,Q)\}}$.
The function ${\cal S}_{NP}(b,Q)$ is parametrized by a 3-parameter 
Gaussian form from a recent global fit to low-energy
Drell-Yan and Tevatron Run-1 $Z^0$ data \cite{Landry:2002ix}.
Combining all the terms, we have:
\begin{eqnarray}
\frac{d\sigma}{dy d q_T^2} 
& = &
\frac{\sigma_0}{S} \ 
\int
\frac{d^2 b}{(2\pi)^2} \ 
e^{-i \vec{q}_T \cdot \vec{b}} \, 
({\cal C}\otimes f)(x_A,b_0/b_*) \,
({\cal C}\otimes f)(x_B,b_0/b_*) \nonumber \\
& \times & 
e^{- {\cal S}_P(b_*,Q) - {\cal S}_{NP}(b,Q) - b^2 \rho(x_A)- b^2 \rho(x_B)} 
+ Y.  
\label{eq:full}
\end{eqnarray}


\begin{figure}[ht] 
\begin{center}
\leavevmode
\vbox{
  \hbox{
     \includegraphics[width=0.45 \hsize]{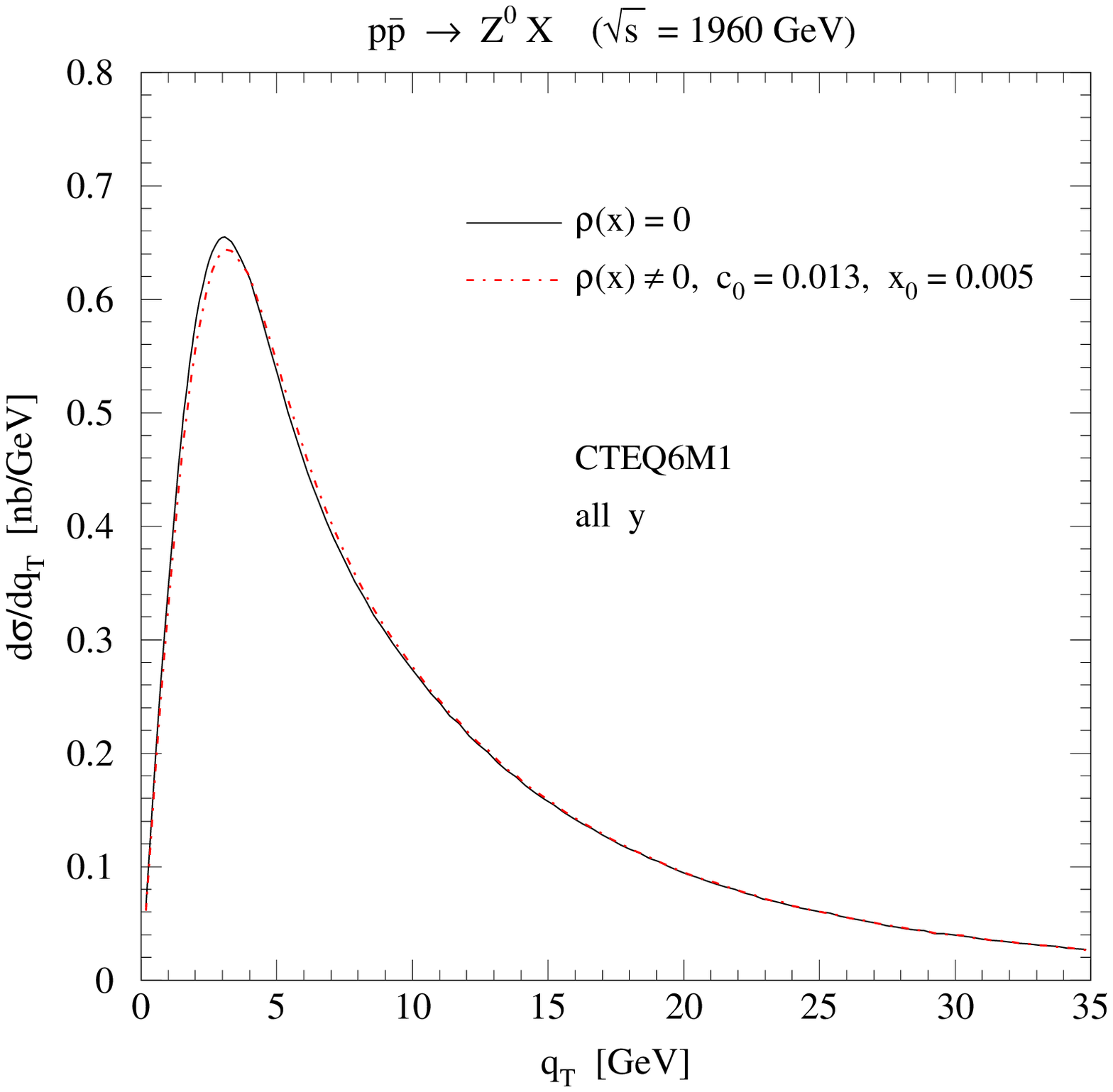} 
     \includegraphics[width=0.45 \hsize]{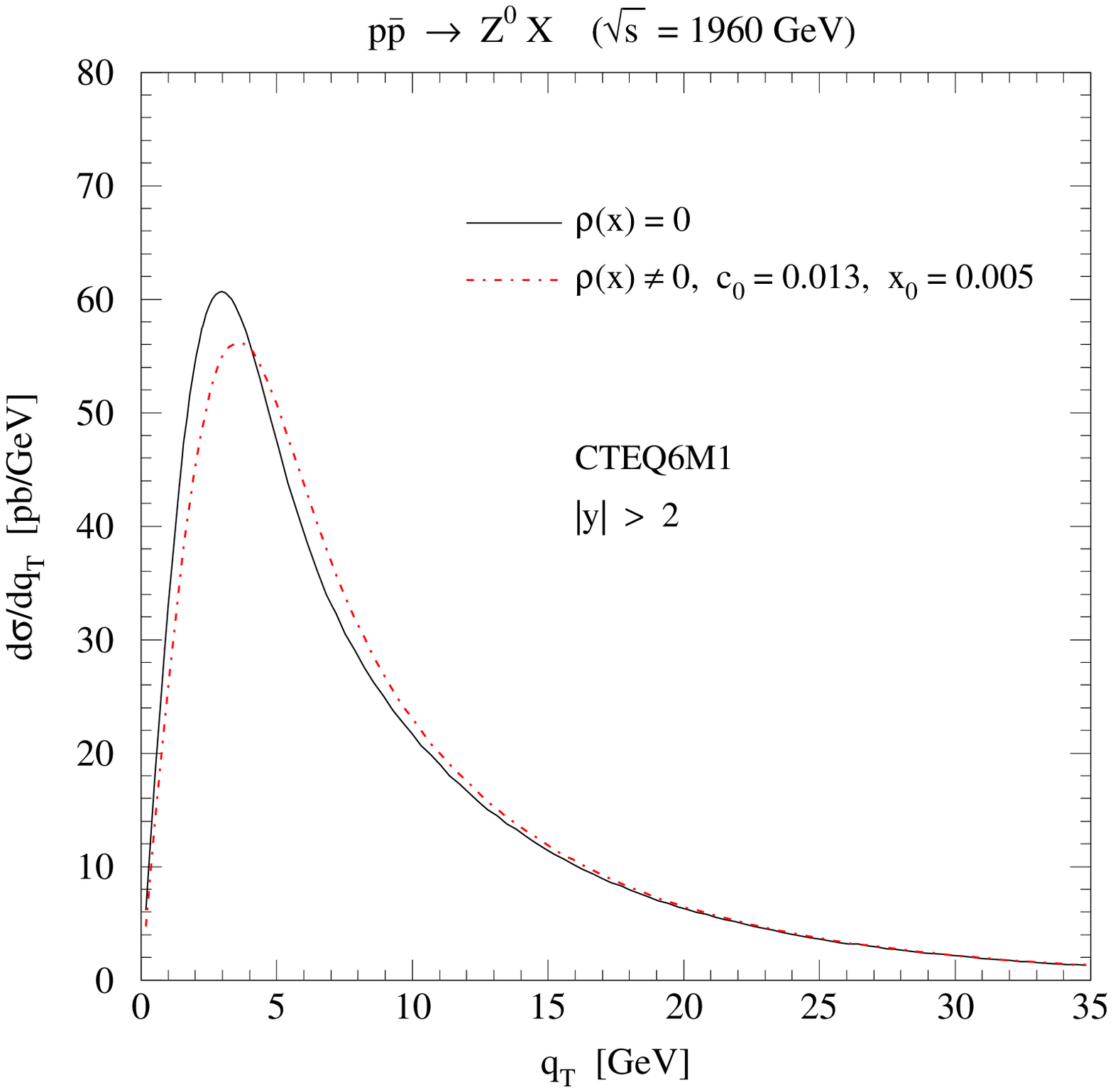} 
  }
}
\\\quad\quad(a) \hspace{6.5cm} (b)
\vskip -00pt
\caption{$q_T$ distributions of $Z^0$ bosons in the  Tevatron Run-2;
(a) integrated over the full range of $Z$ boson rapidities; (b)~integrated
over the forward regions $|y| > 2$.  The solid curve is a standard 
CSS cross section, calculated using the 3-parameter Gaussian 
parametrization~\protect\cite{Landry:2002ix} of the nonperturbative 
Sudakov factor. 
The dashed curve includes additional terms responsible for 
the $q_T$ broadening in the small-$x$ region (cf.~Eq.~(\ref{eq:full})). 
\label{fig:z} 
}
\vskip -00pt
\end{center}
\end{figure}

\begin{figure}[ht] 
\begin{center}
\leavevmode
\vbox{
  \hbox{
     \includegraphics[width=0.45 \hsize]{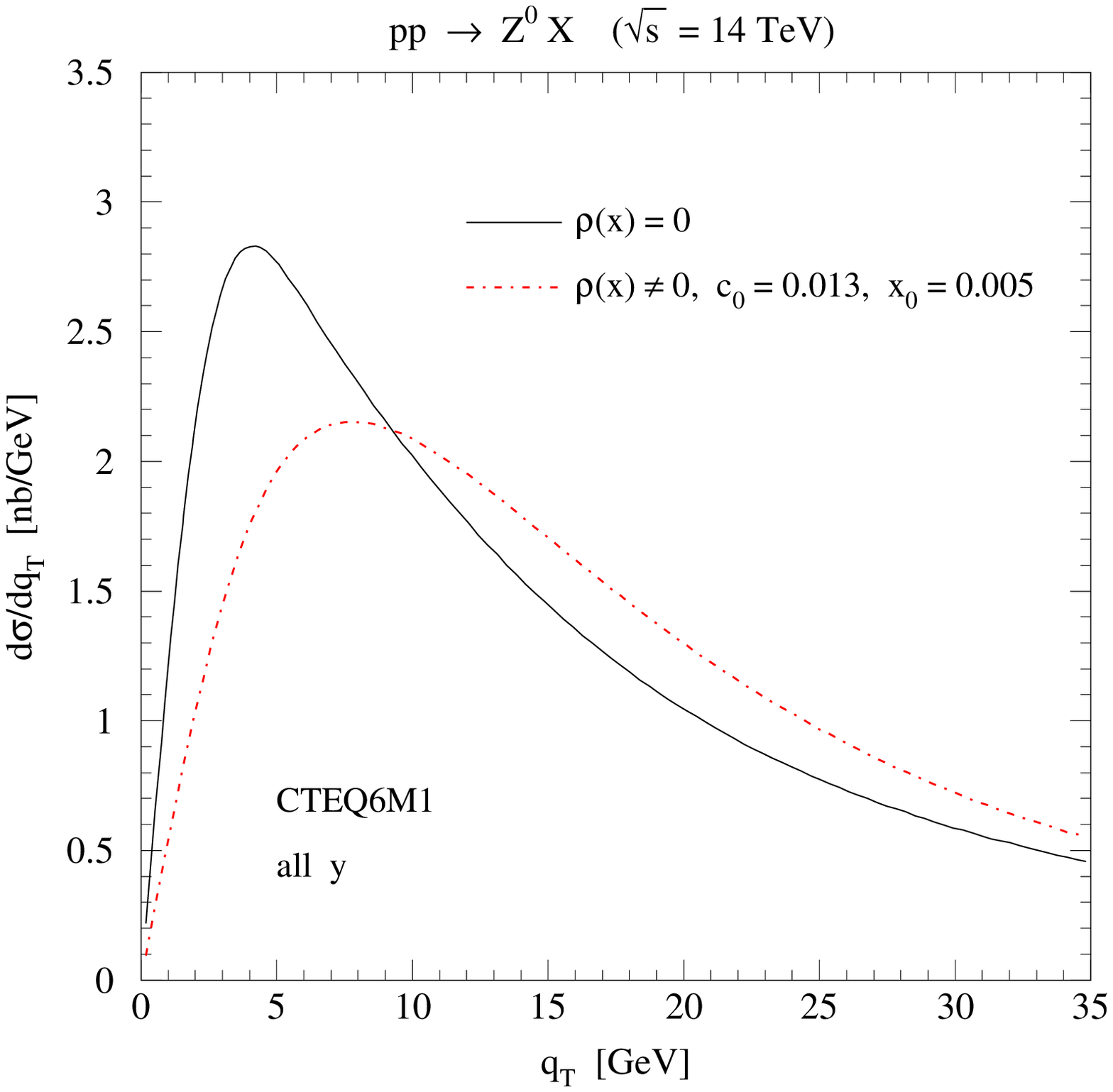}
     \includegraphics[width=0.45 \hsize]{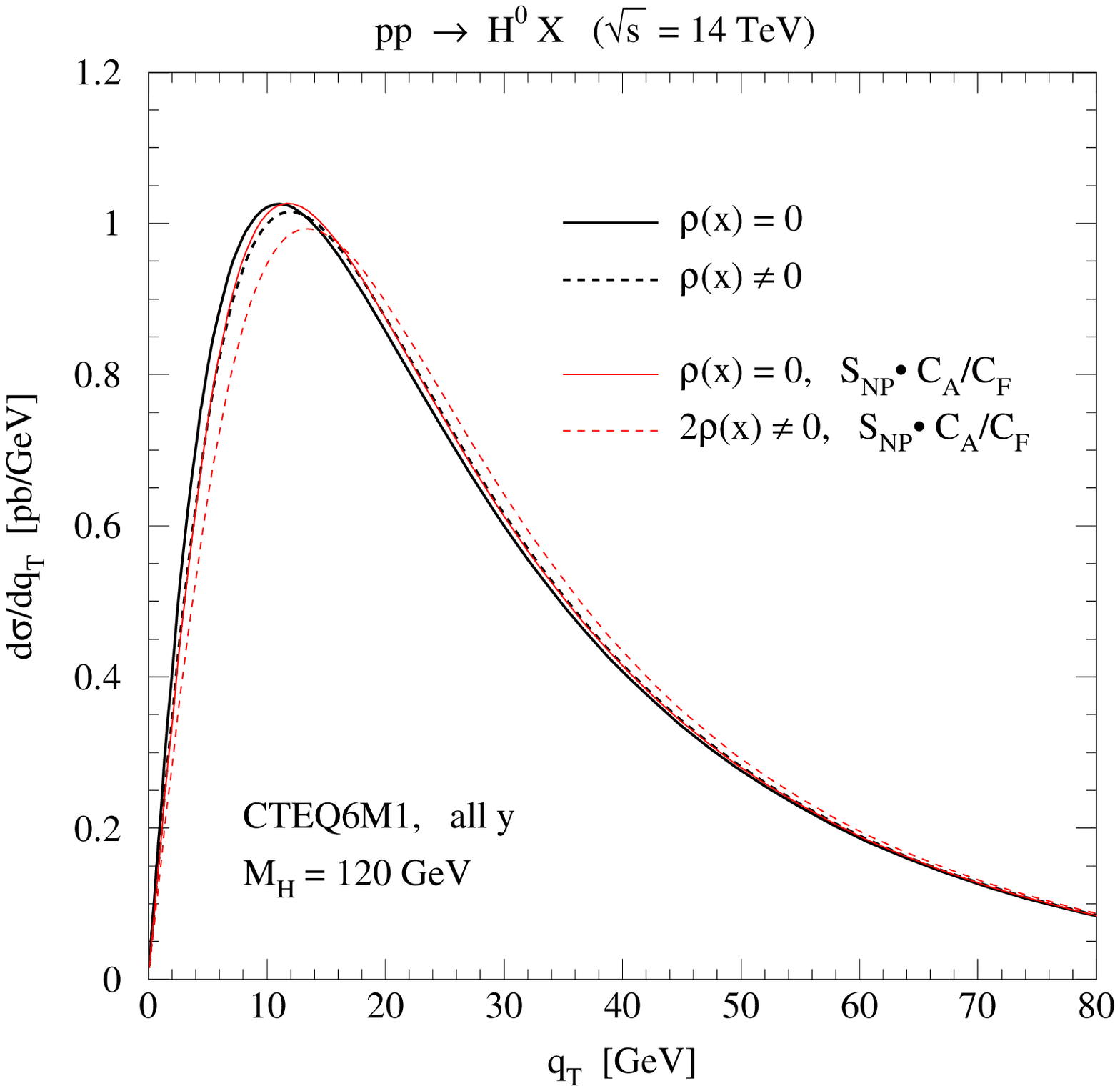}
  }
}
\\\quad\quad(a) \hspace{6.5cm} (b)
\vskip -00pt
\caption{
$q_T$ distributions of (a) $Z^0$ bosons and (b) Standard Model Higgs bosons
at the Large Hadron Collider, integrated over the full range of 
boson rapidities.
\label{fig:h} 
}
\vskip -00pt
\end{center}
\end{figure}

Figs.~\ref{fig:z} and \ref{fig:h} show the comparison of the 
resummed cross section
(\ref{eq:full}) with the additional broadening term 
($\rho(x) \neq 0$) to the resummed cross
section without such a term ($\rho(x) = 0$). We consider cross sections for the
production of $Z^0$ and $H^0$ bosons, calculated according to the procedures
in Refs.~\cite{Balazs:1997xd} and \cite{Balazs:2000wv}, respectively. 
The numerical  calculation was realized using the programs Legacy and ResBos 
\cite{Landry:2002ix,Balazs:1997xd}, and with the CTEQ6M1 parton distribution 
functions~\cite{Stump:2003yu}. The perturbative Sudakov factor was 
included up to ${\cal O}(\alphas^2)$, and the functions 
$({\cal C}\otimes f)$ up to ${\cal O}(\alphas)$. The relevant perturbative
coefficients can be found in Refs.~\cite{Collins:1984kg,ABC}.

Fig.~\ref{fig:z}(a) shows the differential distribution 
$d\sigma/d q_T$ for $Z$ boson production in the Tevatron Run-2,
integrated over the rapidity $y$ of the $Z$ bosons. We observe that 
the distribution (\ref{eq:full}) with the additional small-$x$ term 
(the dashed curve) essentially coincides with the
standard CSS distribution (the solid curve). When $y$ is integrated
over the full range, both resummed cross sections 
are dominated by contributions from $x \sim 0.05 \gg x_0$, where 
the additional broadening (given by the function $\rho(x)$) is negligible. 
For this reason, the Tevatron distributions that are inclusive in $y$ 
(e.g., the Run-1 $Z^0$ boson data) 
will not be able to distinguish the small-$x$ broadening effects from
uncertainties in the nonperturbative Sudakov function ${\cal S}_{NP}$.

In contrast, the small-$x$ broadening does lead to observable differences
in the $q_T$ distributions in the forward rapidity region, where one of 
the initial-state partons carries a smaller momentum fraction than
in the central region. Fig.~\ref{fig:z}(b)
shows the cross section $d\sigma/dq_T$ for $Z$ bosons satisfying $|y|>2$.
The peak of the curve with $\rho(x) \neq 0$ is lower and shifted toward 
higher $q_T$. While this difference 
was not large enough to be observed in the Tevatron Run-1, 
it seems to be measurable in the Run-2 given the improved 
acceptance and higher luminosity of the upgraded Tevatron
collider. The small-$x$ broadening is more pronounced in $W$ boson production 
due to the smaller mass of the $W$ boson.

We now turn to the LHC, where the small-$x$ broadening may be observed in the
whole rapidity range due to the increased center-of-mass energy. 
Fig.~\ref{fig:h}(a) displays the distribution $d\sigma/dq_T$ 
for $Z^0$ production with and without the small-$x$ terms. Here, 
the difference is striking even if $y$ is integrated out. 
Effects of a similar magnitude are present in $W$ boson production,
and they are further enhanced in the forward regions.

The small-$x$ broadening is less spectacular, but visible, 
in the production of light Standard Model Higgs bosons via the effective 
$ggH$ vertex in the limit of a heavy top quark mass. 
Fig.~\ref{fig:h}(b) displays the resummed cross sections for production 
of Higgs bosons with a mass $M_H = 120$ GeV for several choices of 
${\cal S}_{NP}$ and the broadening term. We first compare cross
sections for $\rho(x) = 0$ and $\rho(x) \neq 0$  (thick lines), 
where the functions ${\rho}(x)$ and ${\cal S}_{NP}(b,Q)$ are taken 
to be the same as in $Z^0$ boson production. The difference between 
the two cross sections is not large, due to a harder $q_T$ spectrum 
in the Higgs boson case. The peaking in the $gg$-dominated
$H^0$ distribution occurs at $q_T = 10-20$ GeV, i.e., 
beyond the region where the function $\rho(x)b^2$ play its 
dominant role. This is different from the $q\bar q$-dominated $Z^0$
distribution, where the peak is located at $q_T \sim 5-10$ GeV and
is strongly affected by $\exp{\{-\rho(x)b^2\}}$.
Hence, for the same function $\rho(x)$ as in the $Z^0$ boson case, 
the difference between the curves with and without $\rho(x)$ is 
minimal. 

The harder $q_T$ spectrum in the Higgs boson case is induced by a larger
leading-logarithm coefficient ($C_A$) in $gg$ channels, as compared
to the leading-logarithm coefficient $C_F$ in $q\bar q$ channels. This suggests 
that the $Q$-dependent part (and possibly other terms) 
of the nonperturbative Sudakov function ${\cal S}_{NP}$ 
in Higgs boson production
is also multiplied by a larger color factor than in the Drell-Yan process. 
We estimate this effect by multiplying ${\cal S}_{NP}$ 
by the ratio of the leading color factors 
in Higgs and $Z^0$ boson production processes, $C_A/C_F=9/4$ 
(the thin solid line). 
The resulting change turns out to be small because of the
reduced sensitivity of the Higgs boson cross section to nonperturbative 
contributions.
 
The $\ln(1/x)$ terms may be enhanced in the case of the
Higgs bosons as well, due to the direct coupling of the Higgs bosons 
to gluon ladders. At present, we do not have a reliable estimate
of the small-$x$ broadening in gluon-dominated channels. However, this
broadening would have to be quite large to affect $q_T$ of 10-20 GeV or 
more, i.e., in the region where selection cuts on the $q_T$ of the Higgs boson candidates
will be imposed. For example, increasing the function $\rho(x)$ by a factor 
of two as compared to the $Z^0$ boson case would lead to a distribution 
shown by the thin dashed line. While at $q_T \gtrsim 20$~GeV this effect 
is relatively small as compared to other theory uncertainties 
(e.g., higher-order corrections), it may affect precision calculations of
$q_T$ distributions needed to separate the Higgs boson signal from 
the background in the $\gamma \gamma$ mode.

Additional constraints on the small-$x$ behavior of the resummed 
cross sections in the $gg$ channel could be obtained from examination 
of photon pair production
away from the Higgs signal region. As the mass of the photon 
pair decreases, $\gamma\gamma$ production in the gluon fusion channel via a quark box diagram becomes 
increasingly important. For instance, the subprocess 
$gg \rightarrow \gamma \gamma$ contributes up to 40\% of the total 
cross section at $Q=80$ GeV \cite{Bern:2002jx}. By comparing $q_T$ 
distributions in $p p \to \gamma \gamma$ and  $p p \to Z$ in the same region 
of $Q$, one may be able to separate the $q \bar{q}$ and $gg$ components
of the resummed cross section and learn about the $x$ dependence in
the $gg$ channel.

To summarize, we argue that a measurement of transverse momentum 
distributions of forward $Z$  bosons at the Tevatron will
provide important clues about the physics of QCD factorization and 
possibly discover broadening of $q_T$ distributions associated 
with the transition to small-$x$ hadronic dynamics. 
Based upon the analysis of $q_T$ broadening 
effects observed in semi-inclusive DIS, we have estimated similar effects 
in the (crossed) processes of electroweak boson production at hadron-hadron
colliders. While the estimated impact on the Higgs boson cross section 
$d\sigma/dq_T$ at high $q_T$ was found to be minimal, 
much larger effects may occur 
in $W$ and $Z$ boson production in the forward region at the Tevatron Run-2, 
and at the LHC throughout the full rapidity range. If present,
the small-$x$ broadening will have to be taken into consideration 
in precision studies of electroweak boson production.
Additionally, its observation will provide insights about the
transition to $k_T$-unordered (BFKL-like) dynamics in
multi-scale distributions at hadron-hadron colliders.

\vskip1cm
\noindent
\section*{Acknowledgments}
\indent

 We thank C.~R.~Schmidt for  valuable discussions.
 This work was supported by 
the U.S. Department of Energy under grant  DE-FG03-95ER40908,
the National Science Foundation under grant PHY-0244919, 
and the Lightner-Sams Foundation.


\end{document}